\documentclass[twocolumn,nofootinbib,aps]{revtex4}
\usepackage{graphicx}
\begin{document}

\title{Horizon crossing and inflation with large $\eta$}
\author{William H. Kinney}
\affiliation{Dept.\ of Physics, University at Buffalo, SUNY, Buffalo,
  NY 14260. Email: {\tt whkinney@buffalo.edu}}
 
\begin{abstract} 
I examine the standard formalism of calculating curvature perturbations in inflation at horizon crossing, and derive a general relation which must be satisfied for the horizon crossing formalism to be valid. This relation is satisfied for the usual cases of power-law and slow roll inflation. I then consider a model for which the relation is strongly violated, and the curvature perturbation evolves rapidly on superhorizon scales. This model has Hubble slow roll parameter $\eta = 3$, but predicts a scale-invariant spectrum of density perturbations. I consider the case of hybrid inflation with large $\eta$, and show that such solutions do not solve the ``$\eta$ problem'' in supergravity. These solutions correspond to field evolution which has not yet relaxed to the inflationary attractor solution, and may make possible new, more natural models on the string landscape. 
\end{abstract}

\maketitle  

\section{Introduction}

The physics of density perturbations in inflationary cosmology is well-trodden ground \cite{Mukhanov:1981xt,Hawking:1982my,Starobinsky:1982ee,Guth:1982ec,Bardeen:1983qw,Kodama:1985bj,Mukhanov:1985rz,Mukhanov:1988jd,Sasaki:1986hm,Mukhanov:1990me,Stewart:1993bc,Sasaki:1995aw}. The central ingredients of the ``standard lore'' of inflationary perturbations are (a) the slow roll approximation, and (b) the concept of ``mode freezing'' at horizon crossing, which allows one to relate physical quantities far outside the horizon to variables evaluated when a given length scale was redshifted through the horizon during inflation. While many circumstances have been studied in which the slow roll approximation breaks down \cite{Starobinsky:1992ts,Easther:1995pc,Garcia-Bellido:1996ke,Kinney:1997ne,Wang:1997cw,Seto:1999jc,Adams:2001vc,Inoue:2001zt,Tsamis:2003px}, relatively little attention has been given to the second ingredient of the standard lore: mode freezing at horizon crossing \cite{Grishchuk:1994sj,Martin:1997zd,Grishchuk:1998kp,Martin:1998wd,Unruh:1998ic,Leach:2001zf}. The horizon-crossing formalism must be carefully applied, since mode freezing is a good approximation only for certain  quantities such as the gauge-invariant generalization of the Newtonian potential \cite{Bardeen:1980kt}. Even then decaying-mode solutions exist which are not constant on superhorizon scales. Curvature perturbations can be also generated at late times via a curvaton \cite{Lyth:2001nq,Lyth:2002my}. Other perfectly physical variables can evolve strongly on superhorizon scales and cannot be approximated by their values at horizon crossing.  

In this paper, I examine the horizon-crossing formalism in detail. Section \ref{sec:inflationaryperturbations} gives a review of the general relativistic perturbation theory relevant for inflation, with an emphasis on gauge issues. In Section \ref{sec:horizoncrossing}, I review the horizon-crossing formalism, and derive a relation which must be satisfied in order for the horizon crossing formalism to be valid.  I consider only perturbations which are directly generated from fluctuations in the inflaton.  I show that the horizon-crossing formalism is good for de Sitter space and for power-law inflation, and is valid in an approximate sense for slow roll inflation. In this sense, the slow roll and horizon crossing formalisms are closely related.

In Sec. \ref{sec:ultraslow}, I consider a simple inflation model for which the comoving curvature perturbation ${\cal R}$ evolves rapidly on superhorizon scales, and for which the horizon-crossing formalism is invalid. The model is strongly non-slow roll, with the second Hubble slow roll parameter $\eta = 3$, but nonetheless gives a scale-invariant spectrum of density perturbations. In Sec. \ref{sec:hybrid}, I generalize this result to the case of non-slow roll hybrid inflation considered in Refs. \cite{Garcia-Bellido:1996ke,Kinney:1997ne} and show that, contrary to the conclusions of these papers, inflationary solutions far from slow roll can result in density perturbations consistent with observation. These solutions are identified as early-time transients in which the field evolution has not yet reached the inflationary attractor solution, similar to the case considered by Starobinsky \cite{Starobinsky:1992ts}. In Sec. \ref{sec:modelbuilding}, I comment on the application of these results to string-inspired model building and show that they do {\em not} solve the famous ``$\eta$ problem'' in supergravity. However, non-slow roll evolution opens up a range of largely unexplored dynamical regions of the inflationary parameter space. This may be helpful in constructing more natural models on the string landscape. Section \ref{sec:conclusions} presents a summary and conclusions. 

\section{Curvature perturbations in inflation}
\label{sec:inflationaryperturbations}

In this section, we discuss the evolution of a scalar field dominated cosmology using the useful fluid flow approach \cite{Hawking:1966qi,Ellis:1989jt,Liddle:1993fq,Sasaki:1995aw,Challinor:1998xk}. This approach makes the expression of physical quantities in comoving gauge especially transparent, which is important since the power spectrum of curvature perturbations $P_{\cal R}$ is defined in terms of the intrinsic curvature perturbation on comoving hypersurfaces. Consider a scalar field $\phi$ in an arbitrary background $g_{\mu\nu}$. The stress-energy tensor of the scalar field may be written
\begin{equation}
\label{eq:generalstressenergy}
T_{\mu \nu} = \phi_{,\mu} \phi_{,\nu} - g_{\mu \nu} \left[ {1 \over 2} g^{\alpha\beta} \phi_{,\alpha} \phi_{,\beta} - V\left(\phi\right) \right].
\end{equation}
We can define a fluid four-velocity for the scalar field by
\begin{equation}
\label{eq:deffourvelocity}
u_\mu \equiv {\phi_{,\mu} \over \sqrt{g^{\alpha \beta} \phi_{,\alpha} \phi_{,\beta}}}.
\end{equation}
It is straightforward to show that this vector has unit inner product, $u_\mu u^{\mu} = 1$. We then define the ``time'' derivative of any scalar quantity $f(x)$ by the projection of the derivative along the fluid four-velocity:
\begin{equation}
\label{eq:deftimederiv}
\dot f \equiv u^{\mu} f_{,\mu}.
\end{equation}
In particular, the time derivative of the scalar field itself is
\begin{equation}
\label{eq:defphidot}
\dot\phi \equiv u^{\mu} \phi_{,\mu} = \sqrt{g^{\alpha \beta} \phi_{,\alpha} \phi_{,\beta}}.
\end{equation}
The stress-energy tensor (\ref{eq:generalstressenergy}) in terms of $\dot\phi$ takes the form
\begin{equation}
T_{\mu\nu} = \left[{1 \over 2} \dot\phi^2 + V\left(\phi\right)\right] u_\mu u_\nu + \left[{1 \over 2} \dot\phi^2 - V\left(\phi\right)\right] \left(u_\mu u_\nu - g_{\mu\nu}\right).
\end{equation}
We can then define a generalized density $\rho$ and and pressure $p$ by
\begin{eqnarray}
\label{eq:defrhop}
\rho &&\equiv {1 \over 2} \dot\phi^2 + V\left(\phi\right),\cr
p &&\equiv {1 \over 2} \dot\phi^2 - V\left(\phi\right).
\end{eqnarray}
Note that despite the familiar form of these expressions, they are defined without any assumption of homogeneity of the scalar field or even the imposition of a particular  metric. In terms of the generalized density and pressure, the stress-energy (\ref{eq:generalstressenergy}) is
\begin{equation}
\label{eq:simplestressenergy}
T_{\mu\nu} = \rho u_\mu u_\nu + p h_{\mu \nu},
\end{equation}
where the tensor $h_{\mu\nu}$ is defined as:
\begin{equation}
h_{\mu\nu} \equiv u_\mu u_\nu - g_{\mu\nu}.
\end{equation}
The tensor $h_{\mu\nu}$ can be easily seen to be a projection operator onto hypersurfaces orthogonal to the four-velocity $u_\mu$. For any vector field $A^\mu$, the product $h_{\mu\nu} A^\mu$ is identically orthogonal to the four-velocity:
\begin{equation}
\left(h_{\mu\nu} A^\mu\right) u^\nu = A^\mu \left(h_{\mu\nu} u^\nu\right) = 0.
\end{equation}
Therefore, as in the case of the time derivative, we can define gradients by projecting the derivative onto surfaces orthogonal to the four-velocity
\begin{equation}
\label{eq:defgradient}
\left(\nabla f\right)^\mu \equiv h^{\mu\nu} f_{,\nu}.
\end{equation}
Note that despite its relation to a ``spatial'' gradient, $\nabla f$ is a covariant quantity, {\it i.e.} a four-vector. Identification of $\nabla f$ as a purely spatial gradient is dependent on the choice of gauge. In the case of a scalar field fluid with four-velocity given by Eq. (\ref{eq:deffourvelocity}), the gradient of the field identically vanishes,
\begin{equation}
\left(\nabla \phi\right)^\mu = 0.
\end{equation}

Clearly, the identification of the time derivative (\ref{eq:deftimederiv}) and the gradient (\ref{eq:defgradient}) suggest a favored foliation of the spacetime, {\it i.e.} a choice of gauge. In the case of a scalar field, we can define {\em comoving} gauge as a coordinate system in which spatial gradients of the scalar field $\phi$ are defined to vanish. Therefore the time derivative (\ref{eq:deftimederiv}) is just the derivative with respect to the coordinate time in comoving gauge
\begin{equation}
\dot\phi = \left({\partial \phi \over \partial t}\right)_{\rm c}.
\end{equation}
Similarly, the generalized density and pressure (\ref{eq:defrhop}) are just defined to be those quantities as measured in comoving gauge.

The equations of motion for the fluid can be derived from stress-energy conservation,
\begin{equation}
T^{\mu\nu}{}_{\!;\nu} = 0 = \dot\rho u^\mu + (\nabla p)^\mu + \left(\rho + p\right) \left(\dot u^\mu + u^\mu \Theta\right),
\end{equation}
where the quantity $\Theta$ is defined as the divergence of the four-velocity,
\begin{equation}
\Theta \equiv u^\mu{}_{\!;\mu}.
\end{equation}
We can group the terms multiplied by $u^\mu$ separately, resulting in familiar-looking equations for the generalized density and pressure
\begin{eqnarray}
\dot\rho + \Theta \left(\rho + p\right) = 0&&\cr
(\nabla p)^\mu + \left(\rho + p\right) \dot u^\mu = 0&&.
\end{eqnarray}
The first of these equations, similar to the usual continuity equation in the homogeneous case, can be rewritten using the definitions of the generalized density and pressure (\ref{eq:defrhop}) in terms of the field as
\begin{equation}
\label{eq:generalizedeqofmotion}
\ddot\phi + \Theta \dot\phi + V'\left(\phi\right) = 0.
\end{equation}
This suggests identifying the divergence $\Theta$ as a generalization of the Hubble parameter $H$ in the homogeneous case. In fact, if we take $g_{\mu\nu}$ to be a flat Friedmann-Robertson-Walker (FRW) metric and take comoving gauge, $u^\mu = (1,0,0,0)$,
we have
\begin{equation}
u^\mu{}_{\!;\mu} = 3 H,
\end{equation}
and the generalized equation of motion (\ref{eq:generalizedeqofmotion}) becomes the familiar equation of motion for a homogeneous scalar,
\begin{equation}
\ddot\phi + 3 H \dot\phi + V'\left(\phi\right) = 0.
\end{equation}

Now consider perturbations $\delta g_{\mu\nu}$ about a flat FRW metric,
\begin{equation}
g_{\mu\nu} = a^2\left(\tau\right) \left[\eta_{\mu\nu} + \delta g_{\mu\nu}\right],
\end{equation}
where $\tau$ is the conformal time and $\eta$ is the Minkowski metric $\eta = {\rm diag}\left(1,-1,-1,-1\right)$. We specialize to the case of scalar perturbations, so that the metric perturbations can be written generally in terms of four scalar functions $A$, $B$, ${\cal R}$, and $H_T$:
\begin{eqnarray}
&&\delta g_{00} = 2 A\cr
&&\delta g_{0i} = \partial_i B\cr
&&\delta g_{ij} = 2 \left[{\cal R} \delta_{ij} + \partial_i \partial_j H_T\right].
\end{eqnarray}
If we specialize to comoving gauge, $u^i \equiv 0$, the norm of the four-velocity can be written
\begin{equation}
u^\mu u_\mu = a^2 \left(1 + 2 A\right) \left(u^0\right)^2 = 1,
\end{equation}
and the timelike component of the four-velocity is, to linear order,
\begin{eqnarray}
&&u^0 = {1 \over a} \left(1 - A\right)\cr
&&u_0 = a \left(1 + A\right).
\end{eqnarray}
The velocity divergence $\Theta$ is then
\begin{eqnarray}
\label{eq:thetacomoving}
\Theta &&= u^\mu{}_{\!;\mu} = u^0{}_{\!,0} + \Gamma^\alpha{}_{\!\alpha 0} u^0\cr
&&= 3 H \left[1 - A - {1 \over a H} \left({\partial {\cal R} \over \partial \tau} + {1 \over 3} \partial_i \partial_i {\partial H_T \over \partial \tau}\right)\right],
\end{eqnarray}
where the unperturbed Hubble parameter is defined as
\begin{equation}
H \equiv {1 \over a^2} {\partial a \over \partial \tau}.
\end{equation}
Fourier expanding $H_T$,
\begin{equation}
\partial_i \partial_i H_T = k^2 H_T,
\end{equation}
we see that for long-wavelength modes $k \ll a H$, the last term in Eq. (\ref{eq:thetacomoving}) can be ignored, and the velocity divergence is
\begin{equation}
\label{eq:comovingtheta}
\Theta \simeq 3 H \left[1 - A - {1 \over a H} {\partial {\cal R} \over \partial \tau}\right].
\end{equation}
The metric perturbation ${\cal R}$ is of particular interest, since it determines the intrinsic curvature of comoving hypersurfaces,\footnote{A good explanation of the relationship of this quantity to the gravitational potential and the comoving density perturbation is given in Ref. \cite{Liddle:1993fq}.} 
\begin{equation}
{}^{(3)}R_{\rm c} = 4 \left({k \over a}\right)^2 {\cal R}_{\rm c}.
\end{equation}
We can define the number of e-folds measured relative to the unperturbed metric as integral of the velocity divergence relative to flat hypersurfaces:
\begin{equation}
N \equiv \int{H dt}.
\end{equation}
Similarly, we can define the number of e-folds on a comoving hypersurface as the integral of the velocity divergence along comoving world lines:
\begin{equation}
{\cal N} \equiv {1 \over 3} \int{\Theta d s} = {1 \over 3} \int{\Theta \left[a \left(1 + A\right) d \tau\right]}.
\end{equation}
Using Eq. ({\ref{eq:comovingtheta}) for $\Theta$ and evaluating to linear order in the metric perturbation results in
\begin{equation}
{\cal N} = \int{H d t} - {\cal R},
\end{equation}
so that the comoving curvature perturbation is given by the difference between the number of e-folds ${\cal N}$ on comoving hypersurfaces and the number of e-folds $N$ on flat hypersurfaces:
\begin{equation}
{\cal R} = N - {\cal N}.
\end{equation}
We can express $N$ as a function of the field $\phi$:
\begin{equation}
\label{eq:defN}
N = \int{H d t} = \int{{H \over \dot\phi} d\phi}.
\end{equation}
For monotonic field evolution, we can express $\dot\phi$ as a function of $\phi$, so that
\begin{equation}
{\delta N \over \delta\phi} = {H \over \dot\phi},
\end{equation}
and the curvature perturbation is given by
\begin{equation}
{\cal R} = N - {\cal N} = {\delta N \over \delta\phi} \delta\phi = {H \over \dot\phi} \delta\phi.
\end{equation}
Note that this is an expression for the metric perturbation ${\cal R}$ on comoving hypersurfaces, calculated in terms of quantities defined on {\em flat} hypersurfaces. For $\delta\phi$ produced by quantum fluctuations in inflation, the two-point correlation function is
\begin{equation}
\left\langle \delta\phi^2\right\rangle^{1/2} = {H \over 2 \pi},
\end{equation}
and the curvature perturbation is
\begin{equation}
\label{eq:curvaturepert}
{\cal R} = {H^2 \over 2 \pi \dot\phi},
\end{equation}
which is the well-known result. 

Two comments are in order. First, the comoving perturbation ${\cal R}$ is often referred to as ``gauge invariant'', when it is in fact a manifestly gauge-dependent quantity. On flat hypersurfaces, the intrinsic curvature perturbation is zero! Even when the curvature ${\cal R}$ is defined in a gauge-invariant way, gauge considerations become important when it is evaluated on comoving hypersurfaces. 
This is not a trivial distinction: we will see later that the choice of a preferred foliation of the spacetime is strongly dependent on the late-time behavior of the scalar field. 
Second, it is not immediately obvious from this analysis which values of $\phi$ and $\dot\phi$ to choose when evaluating the perturbation amplitude. The canonical procedure is to evaluate ${\cal R}$ for a given Fourier mode $k$ at ``horizon crossing'', {\it i.e.} when the wavelength of the mode is equal to the horizon size. This is valid as long as the metric perturbation ${\cal R}$ does not evolve when a mode is outside the horizon. (We will see later that the metric perturbation ${\cal R}$ can in some circumstances evolve rapidly on superhorizon scales.) In the next section, we tackle the issue of horizon crossing in detail.

\section{The horizon crossing formalism}
\label{sec:horizoncrossing}

Consider a free scalar field $\varphi$ evolving in an inflationary background. The equation of motion for the field fluctuation can be written in terms of a mode function $u \equiv a \varphi$ in the well-known form:
\begin{equation}
\label{eq:freefieldmode}
{d^2 u_k \over d \tau^2} + \left(k^2 - {1 \over a} {d^2 a \over d\tau^2} \right)
 u_k = 0.
\end{equation}
Let us consider the particularly simple case of a de Sitter background, $H = {\rm const.}$, with conformal time given by
\begin{equation}
\tau = - {1 \over a H}.
\end{equation} 
It is convenient to write the mode equation (\ref{eq:freefieldmode}) in terms of the quantity $y \equiv k / (a H) = - k \tau$, 
\begin{equation}
\label{eq:desittery}
y^2 {d^2 u_k \over d y^2} + (y^2 - 2) u_k = 0,
\end{equation}
with normalized solution
\begin{equation}
\varphi_k = {u_k \over a} = {H \over \sqrt{2 k^3}} \left(1 + i y\right) e^{i y}.
\end{equation}
This is an exact solution of Eq. (\ref{eq:desittery}), valid when the mode is both inside and outside the horizon. There is no necessity to match asymptotic solutions at horizon crossing: we can simply follow the mode from the short wavelength limit to the long wavelength limit. Figure \ref{fig:modefunction} shows the field fluctuation $u_k / a$ in both the long- and short-wavelength limits.
\begin{figure}
\includegraphics[width=3.1in]{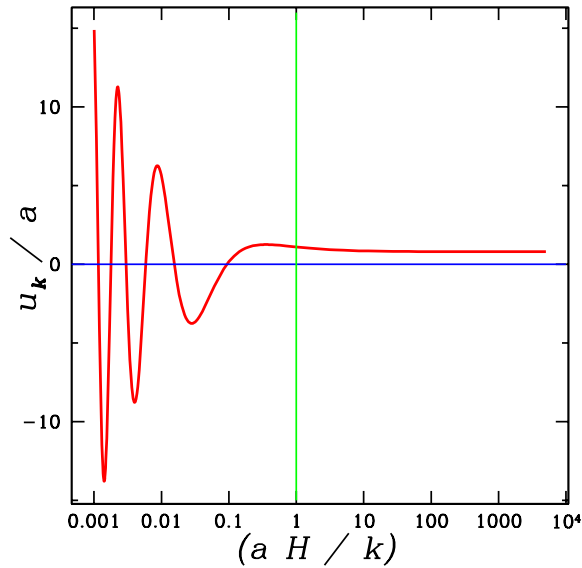}
\caption{\label{fig:modefunction} A free field fluctuation in de Sitter space. The vertical line is horizon crossing, $k = a H$, and the perturbation ``freezes out'' on superhorizon scales. }
\end{figure}

Standard lore has it that quantum modes during inflation ``freeze out'', or cease to evolve, after their wavelength becomes larger than the horizon size. This, of course, depends on which quantity one considers. The amplitude of the mode function $u_k$ does not freeze out at horizon crossing, but instead increases linearly with proper wavelength (or scale factor) outside the horizon,  $u_k \propto a$. It is the field perturbation $\varphi_k \equiv u_k / a$ which becomes constant after horizon crossing.
However, this is only true in the limit of long wavelength. The field fluctuation is not even approximately constant at the time it crosses the horizon: 
\begin{equation}
\left\vert{1 \over \varphi} {d \left(\varphi\right) \over d y}\right\vert_{y = 1} = {1 \over \sqrt{2}}.
\end{equation}
We can compare the amplitude of the field fluctuation at horizon crossing to the amplitude in the asymptotic limit,
\begin{equation}
\left\vert{u_k \over a}\right\vert_{y = 1} = \sqrt{2} \left\vert{u_k \over a}\right\vert_{y \rightarrow 0},
\end{equation}
and the standard expression for the power spectrum of the scalar field corresponds to the asymptotic limit,
\begin{equation}
P_{\varphi}\left(k\right)^{1/2} = \sqrt{k^3 \over 2 \pi^2} \left\vert{u_k \over a}\right\vert_{y \rightarrow 0} = {H \over 2 \pi}. 
\end{equation}
This differs in normalization by a factor of two from its value at horizon crossing. This difference of a constant factor may seem a trivial detail, since asymptotic amplitudes can be easily expressed in terms of amplitudes at horizon crossing by using a constant correction factor. However, this procedure breaks down for modes which exit the horizon but never reach the asymptotic limit \cite{Leach:2000yw}. The situation also becomes more subtle when considering background evolution for which $H \neq {\rm const.}$, especially in the case of curvature fluctuations.

Fluctuations $\delta \phi$ in the inflaton field $\phi$ are more complicated because they couple at linear order to the metric perturbation.  We define a gauge-invariant analog of the field fluctuation $u$ by \cite{Mukhanov:1990me}:
\begin{equation}
u \equiv a \delta\phi - {a \dot \phi \over H} {\cal R}.
\end{equation}
On flat hypersurfaces, the Mukhanov variable $u$ is just the fluctuation in the scalar field, $u = a \delta\phi$.
On comoving hypersurfaces ($\delta\phi \equiv 0$) the variable $u$ is related to the curvature perturbation,
\begin{equation}
{\cal R}_{\rm c} =  \left\vert {H u \over a \dot\phi} \right\vert = \left\vert u \over z\right\vert,
\end{equation}
where the quantity $z$ is defined as
\begin{equation}
\label{eq:defz}
z \equiv {a \dot\phi \over H}.
\end{equation}
For small fluctuations $\delta \phi$, flat and comoving hypersurfaces are related by a gauge transformation.
It is very important for what follows to note that, while the variable $u$ is gauge-invariant, the quantity ${\cal R}$ is gauge-{\em dependent}, and ${\cal R}_{\rm c}$ is defined on spacetime hypersurfaces for which the field fluctuation vanishes, $\delta\phi = 0$. In what follows, we will drop the subscript and implicitly evaluate the metric perturbation ${\cal R}$ in comoving gauge. 
We can write the power spectrum of the curvature perturbation ${\cal R}$ in terms of the Fourier modes $u_k$ as,
\begin{equation}
P_{\cal R}^{1/2}\left(k\right) \equiv \left[{k^3 \over 2 \pi^2} \left\langle {\cal R}^2\right\rangle\right]^{1/2} = \sqrt{k^3 \over 2 \pi^2} \left\vert{u_k \over z}\right\vert,
\end{equation}
where gauge-invariant function $u_k$ satisfies the equation of motion \cite{Mukhanov:1990me}
\begin{equation}
\label{eq:exactmodeequation}
{d^2 u_k \over d\tau^2} + \left(k^2 - {1 \over z} {d^2 z \over d\tau^2} \right)
 u_k = 0.
\end{equation}
The derivative of $z$ is given by:
\begin{equation}
\label{eq:zppoverz}
{1 \over z} {d^2 z \over d\tau^2} = 2 a^2 H^2 \left(1 + \epsilon - {3 \over 2}
 \eta + \epsilon^2 - 2 \epsilon \eta + {1 \over 2} \eta^2 + {1 \over 2}
 \xi^2\right),
\end{equation}
where the Hubble slow roll parameters $\epsilon$, $\eta$ and $\xi^2$ are defined
as derivatives of the background Hubble parameter with respect to the field \cite{Copeland:1993jj,Liddle:1994dx}:
\begin{eqnarray}
\label{eq:defslowrollparams}
\epsilon &&\equiv {m_{\rm Pl}^2 \over 4 \pi} \left({H'(\phi) \over H(\phi)}\right)^2,\cr
\eta &&\equiv {m_{\rm Pl}^2 \over 4 \pi} {H''(\phi) \over H(\phi)},\cr
\xi^2 &&\equiv {m_{\rm Pl}^4 \over 16 \pi^2} {H'(\phi) H'''(\phi) \over H^2(\phi)}.
\end{eqnarray}
This decomposition is valid only for monotonic evolution of the scalar field $\phi$. There is a useful relation between the scale factor and the variable $z$ in terms of the slow roll parameter $\epsilon$ which we will use frequently:
\begin{equation}
z \equiv {a \dot\phi \over H} = - {m_{\rm Pl} \over 2 \sqrt{\pi}} \left(a \sqrt{\epsilon}\right).
\end{equation}
In the case of de Sitter background, $H = {\rm const.}$ and $\dot\phi = 0$, and  the Mukhanov variable becomes identical to a free field fluctuation $u = a \delta\phi$. The equation of motion (\ref{eq:exactmodeequation}) becomes identical to the mode equation for a free field (\ref{eq:freefieldmode}). 

To generalize our discussion to the non-de Sitter case, $H \neq {\rm const.}$, we first consider the exactly solvable case of power-law inflation,
\begin{equation}
a \propto t^{1 / \epsilon},
\end{equation}
where $\epsilon = {\rm const.}$ is the first slow-roll parameter (\ref{eq:defslowrollparams}). In this background, 
\begin{equation}
{1 \over a} {d^2 a \over d \tau^2} = {1 \over z} {d^2 z \over d \tau^2}.
\end{equation}
As in the de Sitter case, the mode equation for curvature perturbations is identical to that for free field perturbations. We again write the mode equation in terms of the variable
\begin{equation}
y \equiv {k \over a H} = {- k \tau \over 1 - \epsilon},
\end{equation}
with the result
\begin{equation}
\label{eq:powerlawmodey}
y^2 \left(1 - \epsilon\right)^2 {d^2 u_k \over d y^2} + \left[y^2 - (2 - \epsilon)\right] u_k = 0.
\end{equation}
The normalized solution is a Hankel function,
\begin{equation}
\label{eq:powerlawmodefunc}
u_k = {1 \over 2} \sqrt{{\pi \over k} \left(y \over 1 - \epsilon\right)} H_\nu\left(y \over 1 - \epsilon\right),
\end{equation}
where
\begin{equation}
\label{eq:powerlawnu}
\nu = {3 - \epsilon \over 2 \left(1 - \epsilon\right)}.
\end{equation}
This equation, like that for the de Sitter case, is exact at all wavelengths, and does not require any matching at horizon crossing ($y = 1$). Likewise, the two-point correlation function for the mode $u_k$ is unambiguous for all wavelengths, and we can consistently define the power spectra for free field and for curvature perturbations in the long-wavelength limit, 
\begin{equation}
\label{eq:freescalarps}
P_\varphi\left(k\right)^{1/2} =  \sqrt{k^3 \over 2 \pi^2} \left\vert{u_k \over a}\right\vert_{y \rightarrow 0},
\end{equation}
and
\begin{equation}
\label{eq:curvatureps}
P_{\cal R}\left(k\right)^{1/2} =  \sqrt{k^3 \over 2 \pi^2} \left\vert{u_k \over z}\right\vert_{y \rightarrow 0}.
\end{equation}
From Eq. (\ref{eq:defz}), we see that the power spectra differ by a constant factor,
\begin{equation}
\label{eq:tensorscalar}
P_\varphi\left(k\right) = {m_{\rm Pl}^2 \over 4 \pi} \epsilon P_{\cal R}\left(k\right).
\end{equation}
Nowhere have we invoked ``mode freezing'' or ``horizon crossing'' when calculating the power spectra. The power spectrum does ``freeze'' in the sense of approaching a time-independent amplitude in the limit $y = k / (a H) \rightarrow 0$, a statement which is true even for  time {\em dependent} $H$. The asymptotic value of the power spectrum is \cite{Stewart:1993bc}:
\begin{equation}
\label{eq:powerlawasym}
\left\vert u_k \over a\right\vert_{y \rightarrow 0} = 2^{\nu - 3/2} {\Gamma\left(\nu\right) \over \Gamma\left(3/2\right)} \left(1 - \epsilon\right)^{\nu - 1/2} \left({H \over k^{3/2} \sqrt{2}}\right) y^{3/2 - \nu}.
\end{equation}
Note that despite the apparent dependence on the quantities $H$ and $y$, this expression is time-independent. This can be derived using the expression \
\begin{equation}
{d H \over d y} = {H \epsilon \over 1 - \epsilon} y^{-1},
\end{equation}
so that
\begin{equation}
\label{eq:derivasym}
{d \over d y} \left(H y^{3/2 - \nu}\right) = 0.
\end{equation}
The spectral index of perturbations can be calculated by taking the derivative of Eq. (\ref{eq:freescalarps}) with respect to $k$ at {\em constant time}, {\it i.e.} $a H = {\rm const.}$,
\begin{equation}
\label{eq:specindexasym}
{d \ln\left(P_\varphi\right) \over d \ln\left(k\right)}\bigg\vert_{a H = {\rm const.}} = 3 - 2 \nu = - {2 \epsilon \over 1 - \epsilon}.
\end{equation}
We can then define a conserved quantity
\begin{equation}
\label{eq:defHstar}
H_*\left(k\right) \equiv H y^{3/2 - \nu}
\end{equation}
which is a time-independent function of wavenumber $k$.
Note, however, that since the function $H_*(k)$ is exactly time-independent, inside and outside the horizon, we are free to evaluate it at any time. Therefore,  it is convenient express $H_*$ in terms of the value of $H$ when the mode crossed the horizon, $y = k / (a H) = 1$:
\begin{equation}
H_*\left(k\right) \equiv H y^{3/2 - \nu} = H\big\vert_{y = 1}.
\end{equation}
We then have an expression for the power spectrum
\begin{equation}
\label{eq:hcps}
P_\varphi^{1/2}\left(k\right) = \sqrt{k^3 \over 2 \pi^2} \left\vert{u_k \over a}\right\vert_{y \rightarrow 0} = A\left(\nu\right) \left({H_*\left(k\right) \over 2 \pi}\right),
\end{equation}
where the constant $A(\nu)$ approaches unity in the de Sitter limit, $\nu = 3/2$:
\begin{equation}
A(\nu) = 2^{\nu - 3/2} {\Gamma\left(\nu\right) \over \Gamma\left(3/2\right)} (\nu - 1/2)^{1/2 - \nu}.
\end{equation}
This is the ``horizon crossing'' formalism. Note that the expression (\ref{eq:hcps}) is {\em not} the value of the power spectrum at horizon crossing, but the value of the power spectrum in the long wavelength limit, expressed in terms of a conserved quantity evaluated at horizon crossing. We can obtain an expression for the spectral index equivalent to (\ref{eq:specindexasym}) by differentiating $H_*\left(k\right)$ with respect to $k$,
\begin{eqnarray}
{d \ln\left(P_\varphi\right) \over d \ln\left(k\right)}\bigg\vert_{k = a H} &=& {d \ln\left[H_*^2\left(k\right)\right] \over d \ln\left(k\right)} \cr 
&=& {a \over H} {d H^2 \over d\left(a H\right)} =  - {2 \epsilon \over 1 - \epsilon}.
\end{eqnarray}
Similarly, we can derive the horizon-crossing expression for the curvature perturbation using Eq. (\ref{eq:tensorscalar}), since the power spectra are simply related by a constant factor,
\begin{eqnarray}
\label{eq:curvasym}
P_{\cal R}^{1/2} &=& \sqrt{k^3 \over 2 \pi^2} \left\vert{u_k \over z}\right\vert_{y \rightarrow 0} = {2 \sqrt{\pi} \over m_{\rm Pl} } {P_\varphi^{1/2} \over \sqrt{\epsilon}}\cr
&=& A\left(\nu\right) {H^2 \over \dot\phi} y^{3/2 - \nu},
\end{eqnarray}
Comparing with Eqs. (\ref{eq:powerlawasym}) and (\ref{eq:derivasym}), we see that there is a conserved quantity relevant to curvature perturbations,
\begin{equation}
\label{eq:Nstarpl}
{d \over d y} \left({H \over \epsilon} y^{3/2 - \nu}\right) = {d \over d y} \left({H^2 \over \dot\phi}y^{3/2 - \nu}\right)  = 0,
\end{equation}
so we can define an analog of $H_*$ for curvature perturbations,
\begin{equation}
\delta N_*\left(k\right) \equiv {H^2 \over 2 \pi \dot\phi}y^{3/2 - \nu} = {H^2 \over 2 \pi \dot\phi}\bigg\vert_{y = 1}.
\end{equation}
The power spectrum for curvature perturbations is
\begin{eqnarray}
P_{\cal R}^{1/2}\left(k\right) = \sqrt{k^3 \over 2 \pi^2} \left\vert{u_k \over z}\right\vert_{y \rightarrow 0} = A\left(\nu\right) \delta N_*\left(k\right).
\end{eqnarray}

We can generalize Eq. (\ref{eq:Nstarpl}) to the case of arbitrary background evolution by noting that
\begin{equation}
{d \over d y} = - {a \dot\phi \over k \left(1 - \epsilon\right)} {d \over d \phi}
= {1 \over y \left(1 - \epsilon\right)} {d \over d N},
\end{equation}
where we have defined $N$ in the conventional way to tend to zero at the end of inflation. This relation is exact for any background evolution. Then we can write the derivative of $\delta N_*$ as, 
\begin{eqnarray}
y {d \over d y} \left({H \over \sqrt{\epsilon}} y^{3/2 - \nu}\right) = && \left({3 \over 2} - \nu\right) \left({H \over \sqrt{\epsilon}} y^{3/2 - \nu}\right)\cr 
&&+  {y^{3/2 - \nu} \over 1 - \epsilon} {d \over d N} \left({H \over \sqrt{\epsilon}}\right). 
\end{eqnarray}
The last term can be evaluated using the flow function relations \cite{Kinney:2002qn}
\begin{equation}
{d H \over d N} = H \epsilon,
\end{equation}
and
\begin{equation}
{d \sqrt{\epsilon} \over d N} =  \sqrt{\epsilon} \left(\eta - \epsilon\right),
\end{equation}
where we have adopted the sign convention
\begin{equation}
\sqrt{\epsilon} = +{m_{\rm Pl} \over 2 \sqrt{\pi}} {H'\left(\phi\right) \over H\left(\phi\right)}.
\end{equation}
We then have an expression valid for arbitrary background evolution,
\begin{equation}
\label{eq:generalNstar}
{d \ln\left(\delta N_*\right) \over d\ln\left(y\right)} = {3 \over 2} - \nu + {2 \epsilon - \eta \over 1 - \epsilon}.
\end{equation}
As expected, this expression vanishes for the de Sitter case, $\nu  = 3/2$, $\epsilon = \eta = 0$, and for the power-law case, for which $\eta = \epsilon$, and $\nu$ is given by Eq. (\ref{eq:powerlawnu}). It is important to emphasize that the relation (\ref{eq:generalNstar}) is a {\em background} relation. The constant $\nu$ is only related to the power spectrum (and therefore the slow roll parameters) in the limit that the field perturbations can be approximated by Hankel functions. 

The generalization to the case of slow roll inflation is straightforward. In the slow roll approximation, $\epsilon$ and $\eta$ are small, independent parameters, and the equation of motion for the curvature perturbation can be written in the approximate form
\begin{equation}
y^2 {d^2 u_k \over d y^2} + \left[y^2 - \left(2 + 2 \epsilon - 3 \eta\right)\right] u_k = 0,
\end{equation}
where we have ignored terms of order $\epsilon^2$, $\eta^2$ and so forth. 
The solution is again a Hankel function:
\begin{eqnarray}
&&u_k \propto y^{1/2} H_\nu\left(y\right),\cr
&&\nu = {3 \over 2} + 2 \epsilon - \eta.
\end{eqnarray}
Unlike the de Sitter and power-law cases, the solution in terms of Hankel functions is only approximate, valid in the limit that the slow roll parameters $\epsilon$ and $\eta$ can be approximated as constant. Likewise, the horizon crossing formalism is only approximate, since for slow roll,
\begin{equation}
\label{eq:Nstarsr}
{d \ln\left(\delta N_*\right) \over d\ln\left(y\right)} = {\epsilon\left(2 \epsilon - \eta\right) \over {1 - \epsilon}} \simeq \epsilon\left(2 \epsilon - \eta\right).
\end{equation}
The variation in  $\delta N_*$ is second order in slow roll, and corrections to observables are expected vanish to first order. This can be verified by considering the asymptotic form of the power spectrum,
\begin{equation}
P_{\cal R}\left(k\right)^{1/2} =  \sqrt{k^3 \over 2 \pi^2} \left\vert{u_k \over z}\right\vert_{y \rightarrow 0}
\propto  \left({H \over \sqrt{\epsilon}}\right) y^{3/2 - \nu}
\end{equation}
We can calculate the spectral index by differentiating this expression with respect to $k$ at constant time, {\it i.e.} $a H = {\rm const.}$,
\begin{eqnarray}
n - 1 &=& {d \ln\left(P_{\cal R}\right) \over d \ln\left(k\right)}\bigg\vert_{a H = {\rm const.}}\cr
&=& 3 - 2\nu = - 4 \epsilon + 2 \eta,
\end{eqnarray}
which is the standard result, but derived without expressing the power spectrum in terms of horizon-crossing variables. We can express the normalization of the power spectrum in terms of horizon crossing variables by using the approximate relation
\begin{equation}
\delta N_* \equiv {H^2 \over 2 \pi \dot\phi} y^{3/2 - \nu} \simeq {H^2 \over 2 \pi \dot\phi}\bigg\vert_{y = 1},
\end{equation}
an approximation which is valid as long as the variation in $\delta N_*$ with scale (\ref{eq:Nstarsr}) is small. It might at first seem that the horizon-crossing formalism is bound to fail in the case of slow roll, since inflation ends when the parameter $\epsilon = 1$. However, for modes already far outside the horizon, the rapid variation in $\delta N_*$ near the end of slow roll inflation will affect all modes identically. Therefore, any effects from the breakdown of slow roll will appear as corrections to the horizon-crossing expressions for the normalization and not the spectral index. Such changes to the normalization of ${\cal R}$ during changes in the cosmological equation of state are well-known \cite{Liddle:1993fq,Deruelle:1995kd}.

So far, everything we have discussed is standard lore. We have, however, emphasized that the horizon crossing formalism, while exact in the de Sitter and power law cases, is only approximate for curvature perturbations in slow roll. In the case of slow roll, the error induced by expressing the power spectrum in horizon crossing variables is of higher order in the slow roll parameters. In the next section, we construct an exactly solvable inflation model for which the horizon crossing approximation is strongly broken, and the standard calculation of even the spectral index gives the wrong answer. 

\section{Ultra-slow roll inflation}
\label{sec:ultraslow}

In this section we discuss the simple case of a scalar field evolving on a constant potential, $V\left(\phi\right) = V_0 = {\rm const.}$, so-called ``ultra-slow roll'' inflation \cite{Tsamis:2003px}. For a constant potential, the equation of motion for the field is independent of the potential:
\begin{equation}
\label{eq:scalareom}
\ddot \phi = - 3 H \dot\phi.
\end{equation}
The equation of motion for a flat Friedmann-Robertson-Walker (FRW) space dominated by the scalar field is
\begin{equation}
H^2 = {8 \pi \over 3 m_{\rm Pl}^2} \left({1 \over 2} \dot\phi^2 + V_0\right).
\end{equation}
This can be written in the equivalent Hamilton-Jacobi formalism 
\cite{grishchuk88,Muslimov:1990be,Salopek:1990jq,Lidsey:1995np},
\begin{equation}
H^2(\phi) - {m_{\rm Pl}^2 \over 12 \pi} \left[H'(\phi)\right]^2 = {8 \pi \over 3 m_{\rm Pl}^2} V_0
\end{equation}
and
\begin{equation}
\dot \phi = - {m_{\rm Pl}^2 \over 4 \pi} H'(\phi).
\end{equation}
We can then solve parametrically for $H$ and $\dot\phi$ as functions of the field value $\phi$. The solution is such that the field comes to a stop at some field value $\phi_0$, which we can set to zero without loss of generality. Taking $\phi > 0$ and $\dot\phi < 0$, the solution to the Hamilton-Jacobi equation is
\begin{equation}
\label{eq:Hus}
H(\phi) = \sqrt{8 \pi V_0 \over 3 m_{\rm Pl}^2} \cosh\left(\sqrt{12 \pi \over m_{\rm Pl}^2} \phi\right).
\end{equation}
We then have
\begin{equation}
\label{eq:phidotus}
\dot\phi = \sqrt{2 V_0} \sinh\left(\sqrt{12 \pi \over m_{\rm Pl}^2} \phi\right),
\end{equation}
which is easily seen to be a solution to the original equation of motion (\ref{eq:scalareom}). As one would expect, the late-time limit of the evolution is de Sitter space, $H = {\rm const.}$ and $\dot\phi = 0$. The slow roll parameters are given by:
\begin{eqnarray}
\epsilon(\phi) &&= 3 \tanh^2\left(\sqrt{12 \pi \over m_{\rm Pl}^2} \phi\right),\cr
\eta(\phi) &&= 3 = {\rm const.},\cr
\xi^2(\phi) &&= 3 \epsilon(\phi).
\end{eqnarray}
Then,
\begin{equation}
{1 \over z} {d^2 z \over d\tau^2} = 2 a^2 H^2 \left(1 - {7\over2} \epsilon + \epsilon^2\right).
\end{equation}
In the early time limit, $\phi \rightarrow \infty$, the parameter $\epsilon \rightarrow 3$ and the equation of state is that of a stiff fluid, $p = \rho$. In the late-time limit, $\phi \rightarrow 0$, the parameter $\epsilon \rightarrow 0$, and the equation of motion for the curvature perturbation approaches
\begin{equation}
\label{eq:desitterlimitmodes}
{d^2 u_k \over d\tau^2} + \left(k^2 - 2 a^2 H^2\right) u_k = 0,
\end{equation}
which is just the mode equation for a field fluctuation in de Sitter space. Note, however, that unlike the exactly de Sitter case, for which $H = {\rm const.}$ and $\epsilon = \eta = \xi^2 = 0$, the second slow roll parameter is always large, $\eta = 3$, even in the limit of nearly exponential expansion, $\dot\phi \rightarrow 0$. The slow roll approximation $\epsilon,\ \eta \ll 1$ is {\em never} valid, no matter how small $\dot\phi$ becomes. Nonetheless the equation of motion for quantum modes (\ref{eq:desitterlimitmodes}) is identical to the equation of motion for quantum modes in an exact de Sitter space, due to the fact that the $\eta$-dependent terms in Eq. (\ref{eq:zppoverz}) cancel for $\eta = 3$. The equations of motion for free field modes and for curvature perturbations are identical, and the solutions are the usual Hankel functions (\ref{eq:powerlawmodefunc}), with $\epsilon \ll 1$, and
\begin{equation}
\nu = \eta - {3 \over 2} = {3 \over 2}.
\end{equation}
The apparent sign inversion in the above expression arises because of the Bunch-Davies boundary condition. The general solution to the mode equation is a a superposition of $H_{\pm 3/2}$ Hankel functions,
\begin{equation}
u_k = A_k H_{3/2}(y) + B_k H_{-3/2}(y)
\end{equation}
The Bunch-Davies boundary condition is the requirement that the mode evolve as a quasi-Minkowski wavefunction at short distance,
\begin{equation}
u_k \propto e^{-i k \tau} \propto e^{i y},
\end{equation}
so that the Bunch-Davies mode function is
\begin{equation}
u_k = N_k H_{3/2}(y).
\end{equation}
Canonical quantization fixes $N_k$, and the mode function is of the form (\ref{eq:powerlawmodefunc}) with $\nu = 3/2$. When $\nu$ changes sign, it merely results in an interchange of the positive- and negative-frequency components of the mode function, which must be accounted for when assigning the Bunch-Davies boundary condition. 

The solution for free field fluctuations is unremarkable. In the asymptotic limit,
\begin{equation}
\left\vert {u_k \over a} \right\vert \propto H y^{3/2 - \nu} = H,
\end{equation}
and we recover a scale-invariant spectrum of free field fluctuations very much like the purely de Sitter case. However, in the late-time limit $\phi \rightarrow 0$, the expression (\ref{eq:curvatureps}) for the spectrum of curvature perturbations diverges,
\begin{equation}
P_{\cal R}^{1/2}\left(k\right) = \sqrt{k^3 \over 2 \pi^2} \left|{u_k \over
 z}\right| = \left({H^2 \over 2 \pi \dot\phi}\right) \rightarrow \infty.
\end{equation}
We can, however, see that the divergence in the $\dot\phi = 0$ limit is a gauge artifact.
In particular, let us examine how we define comoving hypersurfaces in the limit $\dot\phi \rightarrow 0$. We can write the fluid four-velocity $u^\mu$ (\ref{eq:deffourvelocity}) as
\begin{equation}
u^\mu = {\phi^{,\mu} \over \dot \phi},
\end{equation}
which is undefined in the limit of a static field, $\dot\phi = 0$: comoving gauge is singular, and the curvature perturbation on comoving hypersurfaces is a meaningless quantity. It is not possible to define comoving hypersurfaces without field evolution to provide a timelike direction from which to define a foliation of the spacetime. This is not, however, an indication that the spacetime itself is singular, since a zero-curvature foliation of the spacetime is still well-defined. From Eq. (\ref{eq:generalstressenergy}), we can take the stress-energy of a homogeneous, static scalar field to be
\begin{equation}
T_{\mu\nu} =  g_{\mu\nu} V_0,
\end{equation}
with a flat FRW metric, $g_{\mu\nu} = a^2\left(\tau\right) \eta_{\mu\nu}$. Since we are defining the timelike direction to be orthogonal to flat hypersurfaces, quantum fluctuations in the scalar field {\em by construction} do not couple to metric perturbations at linear order. This decoupling of the field perturbation from the curvature in the de Sitter limit can also be seen from the definition of the gauge-invariant mode function $u$ appearing in Eq. (\ref{eq:exactmodeequation}):
\begin{equation}
u = a \delta\phi - {a \dot \phi \over H} {\cal R}.
\end{equation}
In the de Sitter limit, the mode function $u_k$ is independent of the metric perturbation ${\cal R}$,
\begin{equation}
u_k {\rightarrow} a \delta\phi,\ \dot\phi \rightarrow 0,
\end{equation}
and the mode equation (\ref{eq:exactmodeequation}) reduces to that for a free field (\ref{eq:freefieldmode}). More general cases involving the $\dot\phi \rightarrow 0$ limit were considered by Seto, Yokoyama, Kodama, and Inoue \cite{Seto:1999jc,Inoue:2001zt}.
 
We can of course introduce a cutoff in the evolution by forcing an end to inflation at some positive field value $\phi_c$, for example through an instability in an auxiliary field as in hybrid inflation \cite{Linde:1993cn}, or simply by a sudden dropoff from the flat potential for $\phi < \phi_c$. In this case, comoving gauge is well defined, since we can define a constant-time hypersurface at the end of inflation, defined by $\phi = \phi_c$. In this case, the expression (\ref{eq:curvatureps}) for the the power spectrum is finite, and perturbation theory is valid as long as higher-order corrections remain small, which can always be arranged if we take $H$ small enough that $H / \sqrt{\epsilon}$ is small at $\phi = \phi_c$. The power spectrum in the long-wavelength limit is given by the usual expression (\ref{eq:curvasym}), which, for $\nu = 3/2$ and $\phi = \phi_c$ gives:
\begin{equation} 
P_{\cal R}^{1/2}\left(k\right) = \left({H^2 \over 2 \pi \dot\phi}\right)_{\phi = \phi_c}.
\end{equation}
We can calculate the power spectrum by differentiating this expression with respect to $k$ at constant time,
\begin{equation}
n - 1 = {d \ln\left(P_{\cal R}\right) \over d \ln\left(k\right)}\bigg\vert_{a H = {\rm const.}} = 0.
\end{equation}
That is, we have a scale-invariant power spectrum, $n = 1$. It is straightforward to show that the standard horizon-crossing expression for the spectral index gives the wrong answer:
\begin{equation}
{d \ln\left(P_{\cal R}\right) \over d \ln\left(k\right)}\bigg\vert_{k = a H}  = 6,
\end{equation}
as does the usual expression in terms of Hubble slow roll parameters,
\begin{equation}
(n - 1)_{\rm SR} = - 4 \epsilon + 2 \eta = 6.
\end{equation}
The reason for this discrepancy is a breakdown of the horizon crosssing formalism. This can be seen by evaluating Eq. (\ref{eq:generalNstar}) for the case $\epsilon \rightarrow 0$, $\eta = 3$,
\begin{equation}
{d \ln\left(\delta N_*\right) \over d\ln\left(y\right)} = - 3.
\end{equation}
This tells us that the $H / \sqrt{\epsilon}$, and therefore the comoving curvature perturbation, instead of being ``frozen'', is evolving rapidly on superhorizon scales and we cannot express the perturbation in terms of horizon crossing variables. The correct prescription is to evaluate the power spectrum not at horizon crossing, but at the end of inflation,
\begin{equation}
P_{\cal R}^{1/2} = \left({H^2 \over 2 \pi \dot\phi}\right)_{\phi = \phi_c} \neq \left({H^2 \over 2 \pi \dot\phi}\right)_{k = a H}.
\end{equation}
The rapid evolution of the curvature perturbation is a consequence of the rapid divergence of comoving gauge in the limit $\dot\phi \rightarrow 0$. We emphasize that this conclusion is not in conflict with general theorems demonstrating the constancy of gauge-invariant potentials on superhorizon scales. For example, the gauge-invariant variable $u$ is well-behaved on superhorizon scales,
\begin{equation}
u = a \delta\phi - {\dot \phi \over H} {\cal R} \propto a,
\end{equation}
even as the gauge-{\em dependent} quantity $\cal R$ becomes large. This is exactly the ``enhancement'' effect due to the decaying mode considered by Leach {\it et al.} \cite{Leach:2000yw}, expressed in a way that emphasizes the dependence of the result on the late-time behavior of the field. Without an end to inflation, the spacetime smoothly approaches de Sitter space. However, the end of inflation $\phi = \phi_c$ forces the preferred foliation of the spacetime onto comoving hypersurfaces. We then see a scale-invariant spectrum of curvature perturbations with amplitude strongly enhanced relative to its value at horizon crossing. In the next section, we generalize this discussion to the case of hybrid inflation with large $\eta$, a parameter region of great interest for supersymmetric and string-inspired model building.

\section{Hybrid inflation with large $\eta$}
\label{sec:hybrid}

In this section, we generalize the results of Sec. (\ref{sec:ultraslow}) to the case of hybrid inflation \cite{Linde:1993cn} with a potential of the form
\begin{equation}
\label{eq:hybridV}
V\left(\phi\right) = M^4 + {1 \over 2} \mu^2 \phi^2.
\end{equation}
This potential was analyzed in the non-slow roll limit in Refs. \cite{Garcia-Bellido:1996ke,Kinney:1997ne}, which used the horizon crossing formalism to calculate the spectral index of curvature perturbations. We demonstrate that in the limit of large $\eta$, the horizon crossing formalism fails, and the spectral index of curvature fluctuations approaches scale invariance in the limit of $\eta \rightarrow 3$. The equation of motion for the field is:
\begin{equation}
\ddot\phi + 3 H \dot\phi + \mu^2 \phi = 0,
\end{equation}
where for $\dot\phi \ll V'\left(\phi\right)$
\begin{equation}
H \simeq \sqrt{8 \pi M^4 \over 3 m_{\rm Pl}^2} \simeq {\rm const}.
\end{equation}
The equation of motion can be written in terms of $d N = - H dt$ (note that $N$ decreases with increasing time, with $N \equiv 0$ at the end of inflation),
\begin{equation}
{d^2 \phi \over d N^2} - 3 {d \phi \over d N} + \alpha \phi = 0,
\end{equation}
where
\begin{equation}
\label{eq:defalpha}
\alpha \equiv {\mu^2 \over H^2} \simeq {3 \over 8 \pi} \left({\mu^2 m_{\rm Pl}^2 \over M^4}\right).
\end{equation}
The general solution is \cite{Garcia-Bellido:1996ke}:
\begin{eqnarray}
\label{eq:hybfieldevol}
\phi\left(N\right) &=& \phi_+ e^{r_+ N} + \phi_- e^{r_- N},
\end{eqnarray}
where the constants $r_{\pm}$ are given by 
\begin{equation}
\label{eq:rpm}
r_{\pm} \equiv {3 \over 2} \left[1 \mp \sqrt{1 - {4 \over 9} \alpha}\right]. 
\end{equation}
The slow roll parameters are given by \cite{Kinney:1997ne}
\begin{eqnarray}
\epsilon &=& 4 \pi r_{\pm}^2 \left({\phi \over m_{\rm Pl}}\right)^2,\cr
\eta &=& r + \epsilon,
\end{eqnarray}
When the $r_{\pm}$ become imaginary, the field evolution is oscillatory, and no inflation takes place \cite{Garcia-Bellido:1996ke}. We then have the condition for inflation
\begin{equation}
\alpha < 9/4,
\end{equation}
which also ensures that $\epsilon$ is positive-definite.
We will be interested in the limit of small $\phi$, so that
\begin{equation}
\epsilon\left(\phi\right) \simeq  4 \pi \eta^2 \left({\phi \over m_{\rm Pl}}\right)^2  << 1,
\end{equation}
and 
\begin{equation}
\eta \simeq r_{\pm} = {\rm const.}.
\end{equation}
The slow roll limit is $\alpha \ll 1$, for which
\begin{equation}
\label{eq:hybSR}
\eta_{\rm SR} = r_+ \simeq {\alpha \over 3} \ll 1.
\end{equation}
As before, we will take $\phi > 0$ and $\dot\phi < 0$.
The mode equation for curvature perturbations is, neglecting terms of order $\epsilon$ \cite{Kinney:1997ne},
\begin{equation}
y^2 {d^2 u_k \over d y^2} + \left[y^2 - \left(2 + \eta^2 - 3 \eta\right)\right] u_k = 0.
\end{equation}
The solution is again a Bessel function:
\begin{equation}
u_k \propto y^{1/2} H_\nu\left(y\right),
\end{equation}
with index 
\begin{equation}
\label{eq:hybnu}
\nu = \left\vert{3 \over 2} - \eta\right\vert = {3 \over 2} \sqrt{1 - {4 \over 9} \alpha}.
\end{equation}
The absolute value sign in this expression comes from imposition of the Bunch-Davies boundary condition on the mode. The mode equation is identical for $\eta < 3/2$ modes and $\eta > 3/2$ modes, with $\nu \rightarrow 3/2$ in the limit $\eta \rightarrow 3$, which we have already seen recovers a scale-invariant spectrum. A general expression for the curvature power spectrum is obtained in the usual way by taking the long-wavelength limit,
\begin{equation}
P_{\cal R}^{1/2}\left(k\right) \propto \left\vert u_k \over z\right\vert_{y \rightarrow 0} \propto {H \over \sqrt{\epsilon}} y^{r_+}.
\end{equation}
Note in particular that the exponent is
\begin{equation}
{3 \over 2} - \nu = {3 \over 2} \left[1 - \sqrt{1 - {4 \over 9} \alpha}\right] = r_+
\end{equation}
for {\em both} the $\eta = r_+$ and $\eta = r_-$ cases. The spectal index of curvature fluctuations is obtained by taking the constant-time derivative of the asymptotic power spectrum,
\begin{equation}
\label{eq:hybspecindex}
n - 1 = {d \ln\left(P_{\cal R}\right) \over d \ln\left(k\right)}\bigg\vert_{a H = {\rm const.}} = 2 r_+.
\end{equation}
Note that this gives the correct value in the slow roll limit (\ref{eq:hybSR}), but differs substantially for large $\eta$ from the value derived from the horizon-crossing formalism which was used in Refs.  \cite{Garcia-Bellido:1996ke,Kinney:1997ne}. 
Using (\ref{eq:generalNstar}), it is straightforward to show that the horizon-crossing formalism breaks down for large $\eta$:
\begin{equation}
{d \ln\left(\delta N_*\right) \over d\ln\left(y\right)} = {3 /2} - \nu - \eta = r_+ - \eta. 
\end{equation}
For $\eta = r_+$, we have
\begin{equation}
{d \ln\left(\delta N_*\right) \over d\ln\left(y\right)} \simeq 0,
\end{equation}
so the horizon-crossing formalism is good. 
For the $\eta = r_-$ branch, the comoving curvature perturbation evolves rapidly on superhorizon scales,
\begin{equation}
{d \ln\left(\delta N_*\right) \over d\ln\left(y\right)} = r_+ - r_- = - 3 \sqrt{1 - {4 \over 9} \alpha}.
\end{equation}
We therefore recover the standard result in every respect, including the validity of the horizon-crossing formalism, in the slow roll limit. Far from slow roll, however, the standard horizon-crossing expressions are invalid. In the next section, we discuss the details of the dynamics in the context of string-inspired inflation models. 

\section{Application to string-inspired inflation models}
\label{sec:modelbuilding}

In this section, we comment briefly the results of Sec. \ref{sec:hybrid} in the context of model building in string theory and supergravity. It is well known that in string-inspired inflation models, potentials of the form (\ref{eq:hybridV}) arise naturally in many circumstances, most notably in potentials for moduli fields. However, the slow roll limit $\mu \ll M$ is generically unstable to radiative corrections, which forces the mass $\mu$ to be of order the Hubble parameter, 
\begin{equation}
\mu \sim H \sim {M^2 \over m_{\rm Pl}},
\end{equation}
thus spoiling slow roll evolution. This is the famous ``$\eta$ problem'', so called because radiative corrections force $\eta$ to be of order unity \cite{Kachru:2003sx}. With this in mind, we consider scalar field evolution in a potential of the form (\ref{eq:hybridV}). 

Let us write the general solution (\ref{eq:hybfieldevol}) for the field equation in terms of the parameter $\nu$ (\ref{eq:hybnu}):
\begin{eqnarray}
\label{eq:hybfieldgeneral}
\phi\left(N\right) &=& \phi_+ e^{r_+ N} + \phi_- e^{r_- N}\cr
&=& e^{3N/2} \left(\phi_+ e^{-\nu N} + \phi_- e^{\nu N}\right)
\end{eqnarray}
For arbitrary initial conditions, the field will evolve as a superposition of the $\eta = r_+$ and $\eta = r_-$ branches. For example, the case $\mu = 0$ is just ultra-slow roll. In this case $r_+ = 0$ and $r_- = 3$, so that
\begin{equation}
\phi\left(N\right) = \phi_+ + \phi_- e^{3 N}.
\end{equation}
In this limit, we can identify the constant $\phi_+$ as the asymptotic field value $\phi_0$, and the field evolves entirely on the $r_-$ branch. The constant $\phi_-$ can be identified with the end of inflation ($N = 0$), $\phi_- = \phi_c$. The spectral index is given by Eq. (\ref{eq:hybspecindex}), 
\begin{equation}
n - 1 = 2 r_+ = 0,
\end{equation}
in agreement with our earlier analysis. For the more general case of nonzero $\mu$, the $r_+$ branch respresents a slowly-evolving late-time attractor, and the $r_-$ branch is a (relatively) fast-rolling transient which dominates at early times (large $N$). (It is important to note that the solution (\ref{eq:hybfieldgeneral}) is valid only in the limit of small $\phi$, an approximation which breaks down rapidly at early times on the $r_-$ branch.)  The cases $\eta = r_{\pm}$ are just early- and late-time regions of the {\em same} solution for the field evolution, and the spectral index of density fluctuations is the same in both limits:
\begin{equation}
\label{eq:genspecindex}
n - 1 = 2 r_+ = 3 - \sqrt{9 - 4 \alpha},
\end{equation}
where $\alpha = \left(\mu / H\right)^2$  (\ref{eq:defalpha}). This means that transient, non-slow roll behavior can be active during the time when perturbations relevant to cosmology were generated, making available a much richer dynamical phase space for inflationary model building. This may be useful for constructing models on the string ``landscape''. 
 
However, the existence of these non-slow roll solutions does not do much directly to solve the $\eta$ problem. The slow roll limit is obtained for $\alpha \ll 1$, 
\begin{equation}
\label{eq:hybnsr}
(n - 1)_{SR} \simeq {2 \over 3} \alpha = 2 \eta_{\rm SR},
\end{equation}
in agreement with the standard slow roll result. This expression becomes large for $\alpha$ of order unity, hence the $\eta$ problem in inflation. We can use the general equation (\ref{eq:genspecindex}) to constrain $\alpha$. For the WMAP limit $n < 1.32$ \cite{Peiris:2003ff}, the limit on $\alpha$ becomes $\alpha < 0.45$, so that the observational limit on $\mu$ is
\begin{equation}
\mu < 0.67 H.
\end{equation}
The limit from the slow roll expression (\ref{eq:hybnsr}) is nearly identical. If we constrain $n < 1.1$, the corresponding limit is $\mu < 0.39 H$. Therefore, the $\eta$ problem remains unchanged in the presence of the non-slow roll solution, although it is perhaps not as severe as is often supposed.

\section{Conclusions}
\label{sec:conclusions}

We have considered the horizon-crossing formalism for perturbations in inflation. The validity of the horizon crossing formalism depends on the existence of conserved quantities in the cosmological evolution, 
\begin{equation}
{d H_* \over d y} = {d \over d y} \left(H y^{3/2 - \nu}\right) = 0
\end{equation}
for free scalar modes, and 
\begin{equation}
{d \ln\left(\delta N_*\right) \over d\ln\left(y\right)} = {3 \over 2} - \nu + {2 \epsilon - \eta \over 1 - \epsilon} = 0
\end{equation}
for curvature perturbations, where
\begin{equation}
\delta N_* \equiv \left({H^2 \over \dot\phi}y^{3/2 - \nu}\right).
\end{equation}
The constant $\nu$ is the index of Hankel function solution for the mode function $u_k$. 
When either of these conditions is violated, the horizon crossing formalism is invalid for the relevant perturbation. For the cases of de Sitter evolution and power-law inflation, both quantities are exactly conserved, 
\begin{equation}
{d H_* \over d y} = {d \ln\left(\delta N_*\right) \over d\ln\left(y\right)} = 0.
\end{equation}
In the case of slow roll inflation, the conservation law is approximate for curvature perturbations,
\begin{equation}
{d \ln\left(\delta N_*\right) \over d\ln\left(y\right)} = \epsilon\left(2 \epsilon - \eta\right)
\end{equation}
so that corrections to the horizon-crossing expressions are second order in slow roll. 

An example of a circumstance for which the horizon-crossing formalism fails is the simple model of ``ultra-slow roll' inflation, which is a scalar field evolving on an exactly flat potential, $V\left(\phi\right) = {\rm const.}$ In this case, the conservation condition for curvature perturbations is strongly violated,
\begin{equation}
{d \ln\left(\delta N_*\right) \over d\ln\left(y\right)}  = -3,
\end{equation}
indicating that the comoving curvature perturbation ${\cal R}$ is evolving rapidly on superhorizon scales. Curvature perturbations for this model are shown to be scale-invariant, despite the fact that the second Hubble slow-roll parameter is large, $\eta = 3$. The usual horizon-crossing expression for the spectral index gives the wrong answer, $n - 1 = 6$. This can be generalized to the case of hybrid inflation in the limit of large $\eta$, and the dynamics leading to the time-dependence of the curvature perturbation identified as an early-time transient solution in the field evolution. Such transient solutions, despite the strong departure from slow roll, can generate perturbations consistent with observations. Such solutions may be useful for model building on the string landscape, although they do not solve the $\eta$ problem in string theory/supergravity. 

\section*{Acknowledgments}

I thank Richard Woodard and Ali Nayeri for many helpful discussions on the subject of perturbations in the ultra-slow roll case, and Richard Easther for conversations on horizon crossing. I thank Andrew Liddle for comments on a draft version of this paper.

\end{document}